\documentstyle[aps,pre]{revtex}
\begin{document}
\draft
\title{Comment on \\
``Critical behavior of a two-species reaction-diffusion problem'' }
\author{Hans-Karl Janssen}
\address{Institut f\"{u}r Theoretische Physik III, Heinrich-Heine-Universit\"{a}t\\
D\"{u}sseldorf, Germany}
\date{\today}
\maketitle

\begin{abstract}
In a recent paper, de Freitas et al.\ [Phys.\ Rev.\ E {\bf 61}, 6330 (2000)]
presented simulational results for the critical exponents of the two-species
reaction-diffusion system $A+B\rightarrow 2B$ and $B\rightarrow A$ in
dimension $d=1$. In particular, the correlation length exponent was found as
$\nu =2.21(5)$ in contradiction to the exact relation $\nu =2/d$. In this
Comment, the symmetry arguments leading to exact critical exponents for the
universality class of this reaction-diffusion system are concisely
reconsidered.
\end{abstract}


In a recent paper, de Freitas et al.\ \cite{FLSH00} presented a Monte Carlo
study of the two-species reaction-diffusion system $A+B\rightarrow 2B$ and $%
B\rightarrow A$ in dimension $d=1$. They reported the values $\beta
=0.435(10)$ and $\nu =2.21(5)$ for critical exponents of the order parameter
and the correlation length, respectively. The measurement of the short time
scaling exponent $\theta ^{\prime }$ \cite{JSS89} seems consistent with the
scaling laws $\theta ^{\prime }=-\eta /z$ and $2\beta =\nu (d+\eta )$ using
the exact value $z=2$ of the dynamical exponent. The critical
reaction-diffusion system above \cite{WOH98} belongs in general to the
universality class of directed percolation (DP) processes coupled to a
secondary conserved density that I call DP-C \cite{Ja(u)} in analogy to the
Model C. In the same manner as the universal behavior of the critical
dynamics of a relaxing non-conserved order parameter near equilibrium (Model
A) is changed to Model C by the coupling to a conserved density, DP
processes are changed to DP-C processes. de Freitas et al.\ assume equal
diffusion constants for both species. Therefore a special DP-C process
called KSS and identified by Kree et al. \cite{KSS89} several years ago
describes the special system studied here. In the appendix of their paper,
Kree et al.\ show by means of a Ward identity that the correlation length
exponent obeys the exact relation $\nu =2/d$. This yields $\nu =2$ in one
dimension. Thus the value of $\nu $ reported by de Freitas et al.\ cannot be
correct. Likewise their conjectured simple fractions for the critical
exponents have to be rejected.

Because the arguments leading to exact critical exponents of the DP-C
processes are more or less implicit in several papers \cite{WOH98,KSS89}, I
will reconsider their derivation in this comment, and show that they all
have their roots in particular symmetry properties.

The Langevin dynamics of the DP-C class can be described by the dynamic
functional \cite{Ja76,DeDo76}
\begin{eqnarray}
{\cal J} &=&\int dt\,d^{d}x\,\biggl\{\widetilde{s}\Bigl[\partial
_{t}+\lambda (\tau -\nabla ^{2}+fc)+\frac{\lambda }{2}(gs-\widetilde{g}%
\widetilde{s})\Bigr]s  \nonumber \\
&&+\widetilde{c}\Bigl[\partial _{t}c-\gamma \nabla ^{2}(c+\sigma s)\Bigr%
]-\gamma (\nabla \widetilde{c})^{2}\biggr\}\ .  \label{DynFunkt}
\end{eqnarray}
Here $s$ and $c$ are the densities of the percolating agent and the
conserved field, respectively. In the case of the reaction-diffusion system
above, $s\propto n_{B}$ and $c\propto n_{A}+n_{B}$, where $n_{A}$ and $n_{B}$
denote the densities of the $A$ and $B$ particles, respectively. The
conjugated response fields are denoted by $\widetilde{s}$ and $\widetilde{c}%
. $ Stability requires $g>2\sigma f$. Green functions (correlation and
response functions) are obtained by integrating the fields against a weight
factor $\exp (-{\cal J})$.

The functional ${\cal J}$, Eqn.\ (\ref{DynFunkt}), possesses the following
symmetries under three transformations involving a constant continuous
parameter $\alpha $:
\begin{eqnarray}
\text{I:}\quad &&\widetilde{c}\rightarrow \widetilde{c}+\alpha \ ;  \label{I}
\\
\text{II:}\quad &&c\rightarrow c+\alpha \ ,\quad \tau \rightarrow \tau
+f\alpha \ ;  \label{II} \\
\text{III:}\quad &&s\rightarrow \alpha s\ ,\quad \widetilde{s}\rightarrow
\alpha ^{-1}\widetilde{s}\ ,\quad \sigma \rightarrow \alpha ^{-1}\sigma \
,\quad g\rightarrow \alpha ^{-1}g\ ,\quad \widetilde{g}\rightarrow \alpha
\widetilde{g}\ .  \label{III}
\end{eqnarray}
Moreover, ${\cal J}$ is invariant under the inversion:
\begin{equation}
\text{IV:}\quad \widetilde{c}\rightarrow -\widetilde{c}\ ,\quad c\rightarrow
-c\ ,\quad \sigma \rightarrow -\sigma \ ,\quad f\rightarrow -f\ .  \label{IV}
\end{equation}
In the particular case $\sigma =0$, the time inversion
\begin{eqnarray}
\text{V:}\quad &&\sqrt{g/\widetilde{g}}\,s({\bf x},t)\leftrightarrow -\sqrt{%
\widetilde{g}/g}\,\widetilde{s}({\bf x},-t)\ ,\quad  \nonumber \\
&&c({\bf x},t)\rightarrow c({\bf x},-t)\ ,\quad \widetilde{c}({\bf x}%
,t)\rightarrow c({\bf x},-t)-\widetilde{c}({\bf x},-t)  \label{V}
\end{eqnarray}
yields a further discrete symmetry transformation. The symmetry V
distinguishes the KSS from general DP-C processes.

Symmetry I results from the conservation property of the field $c$.
Symmetries III and IV show that dimensionless invariant coupling constants
and parameters are defined by $u=\widetilde{g}g\mu ^{-\varepsilon }$, $%
v=f^{2}\mu ^{-\varepsilon }$, $w=\sigma \widetilde{g}f\mu ^{-\varepsilon }$,
and the ratio of the kinetic coefficients $\rho =\gamma /\lambda $. Here $%
\mu ^{-1}$ is a convenient mesoscopic length scale and $\varepsilon =4-d$.

Dimensional analysis and the scaling symmetry III applied to the Green
functions $G_{N,\widetilde{N};M,\widetilde{M}}=\langle \lbrack s]^{N}[%
\widetilde{s}]^{\widetilde{N}}[c]^{M}[\widetilde{c}]^{\widetilde{M}}\rangle $
gives
\begin{eqnarray}
G_{N,\widetilde{N};M,\widetilde{M}} &=&\alpha ^{\widetilde{N}-N}G_{N,%
\widetilde{N};M,\widetilde{M}}(\{{\bf x},t\},\tau ,\alpha ^{-1}\sigma
,\alpha \widetilde{g},\alpha ^{-1}g,f,\lambda ,\gamma ,\mu ) \\
&=&\sigma ^{\widetilde{N}-N}F_{N,\widetilde{N};M,\widetilde{M}}(\{\mu {\bf x}%
,\gamma \mu ^{2}t\},\mu ^{-2}\tau ,u,v,w,\rho )  \label{SkalSig} \\
&=&\bigl(g/\widetilde{g}\bigr)^{(\widetilde{N}-N)/2}F_{N,\widetilde{N};M,%
\widetilde{M}}^{\prime }(\{\mu {\bf x},\gamma \mu ^{2}t\},\mu ^{-2}\tau
,u,v,w,\rho )\ ,  \label{Skalg}
\end{eqnarray}
where it is assumed that $\sigma \geq 0$ for convenience. Of course, Eqn.\ (%
\ref{SkalSig}) cannot be used if $\sigma =0$.

The critical scaling properties of the Green functions can be extracted from
the invariant functions $F$ and $F^{\prime }$ by applying the
renormalization group. To extract UV-finite quantities from the field theory
one introduces bare fields and parameters and renormalizes by appropriate $Z$%
-factors. For example one uses the scheme $s\rightarrow \mathaccent"7017{s}%
=Z_{s}^{1/2}s$, $\widetilde{s}\rightarrow \mathaccent"7017{\tilde{s}}=Z_{%
\tilde{s}}^{1/2}\widetilde{s}$, $\tau \rightarrow \mathaccent"7017{\tau }%
=Z_{\tau }\tau +\mathaccent"7017{\tau }_{c},$, $f\rightarrow \mathaccent"7017%
{f}=Z_{f}v^{1/2}\mu ^{\varepsilon /2}$, etc. Here $\mathaccent"7017{\tau }%
_{c}$ denotes the critical value of $\mathaccent"7017{\tau }.$ The $Z$%
-factors have to absorb all the UV-infinities (the $\varepsilon $-poles in
dimensional regularization). They can only depend on the invariant
parameters $u$, $v$, $w$, and $\rho $.

The objects of the calculation are the vertex functions $\Gamma _{\widetilde{%
N},N;\widetilde{M},M}$, i.e., the one-particle irreducible amputated
diagrams with $\widetilde{N}$ $\widetilde{s}$-legs, $N$ $s$-legs, $%
\widetilde{M}$ $\widetilde{c}$-legs, and $M$ $c$-legs. It is easily seen
that diagrams with loops do not contribute to vertex functions with $%
\widetilde{M}\geq 1$. Thus, these vertex functions are trivial and given by
the corresponding terms displayed in the dynamic functional ${\cal J}$,
Eqn.\ (\ref{DynFunkt}). Hence, the renormalizations are trivial:
\begin{equation}
\mathaccent"7017{\widetilde{c}}=\widetilde{c},\quad \mathaccent"7017{c}%
=c,\quad \mathaccent"7017{\gamma }=\gamma ,\quad \mathaccent"7017{\sigma }%
\mathaccent"7017{s}=\sigma s\ .  \label{trivRen}
\end{equation}

Symmetry II in connection with the trivial renormalization of $c$, Eqn.\ (%
\ref{trivRen}), shows that $f$ is renormalized with the same $Z$-factor as $%
\tau $: $Z_{f}=Z_{\tau }$. It follows the simple relation
\begin{equation}
\frac{\tau }{\mu ^{\varepsilon /2}}=\frac{(\mathaccent"7017{\tau }-%
\mathaccent"7017{\tau }_{c})}{\mathaccent"7017{f}}v^{1/2}\ .  \label{SimpRel}
\end{equation}
At a fixed point $v_{\ast }$ different from $0$ and $\infty $, $\tau $
changes according to this relation by a change of the momentum scale $\mu
\rightarrow \mu l$ (holding bare parameters fixed) as $\tau \rightarrow \tau
(l)=\tau l^{\varepsilon /2}$. Thus, one finds from the Eqns.\ (\ref{SkalSig},%
\ref{Skalg}) the scaling properties of the Green functions at a fixed point
with finite values for $u_{\ast }$, $v_{\ast }$, $w_{\ast }$, and $\rho
_{\ast }$ different from $0$ and $\infty $ (the existence of such fixed
points can be demonstrated in the $\varepsilon $ expansion \cite
{WOH98,Ja(u),KSS89})
\begin{equation}
G(\{{\bf x},t\},\tau )=l^{\delta _{G}}G(\{\mu l{\bf x},\gamma \mu
^{2}l^{2}t\},\mu ^{-2}\tau /l^{2-\varepsilon /2})\ .  \label{GSkal}
\end{equation}
In Eqn.\ (\ref{GSkal}) $\delta _{G}=(N+\widetilde{N}+M+\widetilde{M}%
)d/2+(N\eta +\widetilde{N}\widetilde{\eta })/2$ denotes the scaling exponent
of $G$. $\delta _{G}$ combines the normal and anomalous dimensions of the
fields involved in the Green function $G$. From Eqn.\ (\ref{GSkal}) one can
gather the exact values of the dynamical exponent $z=2$ and the correlation
length exponent $\nu =1/(2-\varepsilon /2)=2/d$.

The renormalization of quantities invariant under the transformation
defining symmetry III, like $F$ or $F^{\prime }$, involve only the product
of the field renormalizations $Z=Z_{\tilde{s}}Z_{s}$. Thus, one has a
freedom to define one of this factors. With respect to Eqns.\ (\ref{SkalSig}%
) and (\ref{trivRen}) it is convenient to choose the trivial renormalization
$\mathaccent"7017{\sigma }=\sigma $ together with $Z_{s}=1$. Then the Green
functions $G$ have the same scaling properties under renormalization as the
invariant functions $F$. One could also define $Z_{s}\neq 1$ and renormalize
$\mathaccent"7017{\sigma }=Z_{s}^{-1/2}\sigma $. The renormalization
properties of $F$ are not affected by this choice. Hence one finds the same
critical scaling properties of the correlation and response functions as for
the previous one. It follows the anomalous dimension of the field $s$ as $%
\eta =0$. Then the anomalous dimension $\widetilde{\eta }$ of the response
field $\widetilde{s}$ is given by the logarithmic derivative of $Z$ with
respect to the momentum scale $\mu $ at the fixed point. $\widetilde{\eta }$
is the only scaling exponent that one has to determine by perturbation
theory.

For the KSS processes $\sigma $ vanishes and one cannot follow the strategy
of the last paragraph. However, for the KSS processes the time inversion
symmetry V can be explored. With respect to this symmetry and Eqn.\ (\ref
{Skalg}) it is now convenient to choose the ratio $\mathaccent"7017{\tilde{g}%
}/\mathaccent"7017{g}=\widetilde{g}/g$ trivially renormalized together with $%
Z_{s}=Z_{\tilde{s}}=Z^{1/2}$. Then the Green functions $G$ have the same
critical scaling as the invariant functions $F^{\prime }$. The logarithmic
derivative of $Z$ at the fixed point yields in this case the value of $\eta =%
\widetilde{\eta }$.

In summary, the DP-C processes offer exact relations for some critical
exponents including $z=2$ and $\nu =2/d$. In the KSS case one has $\eta =%
\widetilde{\eta }$ as in DP, but with another value. In the more
general case with $\sigma \neq 0$ (and $\sigma f<0)$ one finds
$\eta =0$, which yields via the relation $\beta =\nu (d+\eta )/2$
the exact order parameter exponent $\beta =1$. The discrepancy
with the exact and the simulational result for $\nu $ by roughly
$10\%$ may have the origin in corrections to scaling. It is
therefore desirable that the authors reconsider their simulations
and provide a careful analysis of such corrections.

\end{document}